# A Distributed Feedback System for Rapid Stabilization of Arbitrary Process Variables[*]

B. Bevins and A. Hofler, Jefferson Lab, Newport News, VA 23606, USA

## Abstract

In large process control systems it frequently becomes desirable to establish feedback relationships that were not anticipated during the design phase of the project. The "Generic Lock" architecture discussed in this paper makes it possible for system operators to implement new feedback loops between arbitrary process variables quickly and with no disturbance to the underlying control system. Any available process variables may be selected for the input and output of the loops so created, regardless of their physical or logical separation. The system allows multiple user interface points distributed through the control system while ensuring consistency among the feedback loops. This system can be used to quickly prototype and test new control philosophies or to control temporary hardware arrangements without any new software development. It is implemented at the Thomas Jefferson National Accelerator Facility using the Common Device (CDEV) [1] framework on top of the Experimental Physics and Industrial Control System (EPICS) [2]. This paper discusses the architecture, implementation, and early usage of the system.

## 1 INTRODUCTION

One of the fundamental entities in any digital process control system is the discrete closed-loop feedback controller. Though many such controllers are typically designed into a large system, it is difficult to anticipate all possible useful feedback relationships. Hardware details, operating modes, and even control philosophies may evolve over time to meet changing needs. Control systems must have the flexibility to evolve with them.

A Generic Lock Server has been developed at Jefferson Lab that enables on the fly creation and configuration of feedback loops using any available control system I/O signals. With this system, feedback loops can be created as prototypes for new control ideas or to satisfy temporary operational requirements with absolutely no disruption of the existing control system and no new programming effort required. It builds upon previous work at the lab in software server based process control [3]. At present only single-input, single-output (SISO) loops are available, but multiple-input, multiple-output (MIMO) loops will be added in the near future.

## 2 DESIGN

### 2.1 Background

The control system software for the CEBAF accelerator at Jefferson Lab is implemented in two layers: EPICS databases running on the input/output controllers (IOC's) and programs running on Unix hosts that communicate with the IOC's by Channel Access (CA). Among the latter are programs that implement feedback loops to stabilize various machine parameters against disturbances at < 1Hz [4]. They are referred to as "Slow Locks" to distinguish them from the Fast Feedback System, which uses dedicated hardware to achieve much faster sampling rates [5]. They come in various distinct flavors: orbit locks, energy locks, current locks, and helicity-correlated asymmetry locks. Some calculate their feedback gains with data from on-line accelerator model servers. Others empirically measure their response functions during an explicit calibration step [6]. Still others use operator-entered gains that have been calculated off-line or optimized interactively. Though they share some code, each lock type exists in its own dedicated server with its own GUI. Adding new lock types has been difficult and time consuming.

### 2.2 The Generic Lock Server

An effort is now underway to unify the various lock types into a common architecture that will be more easily maintainable and extensible. The first fruit of this effort is the Generic Lock Server. This server can host multiple lock types simultaneously and allows the interactive creation and destruction of new locks at runtime. It stores configuration information in a human readable Extensible Markup Language (XML) file. The Generic Lock Server presently handles a new class of general purpose Proportional, Integral, Derivative controller (PID) locks as well as the specialized current and asymmetry locks.

### 2.3 The PID Locks

The functionality of the PID locks is derived from the EPICS CPID record, extended to allow both the input and output signals to be arbitrarily specified at runtime. Any variable in the control system accessible through CDEV can be used. This includes all CA signals as well as

---

[*] Work supported by the U. S. Department of Energy, contract DE-AC05-84-ER40150

signals from other CDEV servers, including model servers and other locks.

A GUI allows users to control all the PID locks and create new ones from anywhere on the network. Once a new lock has been created, the user enters names for input and output of the feedback loop, PID gain parameters, and output signal limits. The lock can then be activated.

## 3 IMPLEMENTATION

### 3.1 Server Architecture

The lock server programs are implemented in C++ as CDEV Generic Servers [7] where each lock is a virtual device that exposes a set of virtual attributes containing the lock's operating parameters. The Generic Lock Server uses this same setup, but takes advantage of more recent developments in C++. It uses the containers and strings of the C++ standard library (formerly the standard template library) to avoid memory management issues. This has resulted in a very robust product being developed in much less time that would otherwise be possible.

Figure 1 shows the relationships among the main classes used by the Generic Lock Server. In order to minimize coupling among the components, the LockServer engine uses a LockFactory class to construct locks of various types. The abstract Lock class maintains a private list of all the locks that have been instantiated and controls access to them. Each derived lock type registers itself with the factory at initialization. All that is necessary to add a new lock type is to relink the server with the new object file for the lock type. The ConfigParser and LockFactory classes are singletons.

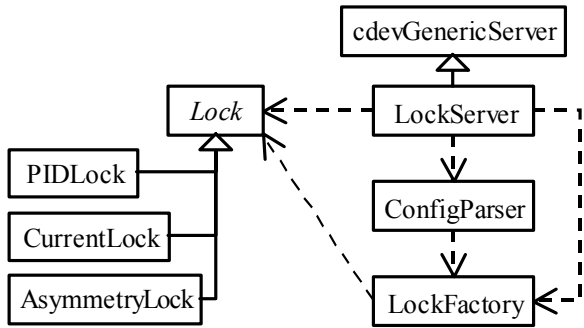

**Figure 1: Simplified Lock Server Class Diagram**

### 3.2 Configuration Files

The XML configuration file is both parsed and written using the "non-commercial" Qt/X11 toolkit from Trolltech [8]. The Qt XML Document Object Model (DOM) interface is encapsulated in the ConfigParser class used by the LockServer. A partial XML configuration file is shown below.

Within the main <lockConfig> element, there are zero or more <Lock> elements and zero or more <device> elements. Each <Lock> triggers the construction of a new lock of the specified type with the specified name. Each <device> contains zero or more <attribute> elements. The parser attempts to map the names of each <device>/<attribute> pair into a CDEV device/attribute existing in the server and set its value accordingly.

```
<lockConfig>
 <Lock name="PIDLock02" type="PIDLock" />
 <device name="PIDLock02" >
  <attribute name="GainD" value="0" />
  <attribute name="GainI" value="1" />
  <attribute name="GainP" value="0" />
  <attribute name="InputName"
       value="ILI1L_PHASEerror"/>
  <attribute name="Interval" value="4" />
  <attribute name="MaxChange" value="0.1" />
  <attribute name="MaxPos" value="25" />
  <attribute name="MinPos" value="15" />
  <attribute name="OutputName" value="R1XXPSET" />
  <attribute name="SetPoint" value="0" />
  <attribute name="Description" value="North Linac First
       Pass Gang Phase" />
 </device>
<!--More locks of various types would follow here. ->
</lockconfig>
```

### 3.3 User Interface

The GUI's for the slow locks are scripted in Tcl/Tk. The PID Lock GUI is shown in Figure 2. All the PID locks are presented in a scrollable list and each one can be collapsed or expanded using the arrow buttons on the left. Output values from the locks are in the darker boxes while user inputs are in the lighter boxes. The button to the left of each lock's textual description field turns the lock on and off. A lock can be deleted using the button in the upper right corner of its display. A new lock is created using the button in the far lower left. The new lock is assigned the next available device name, first reusing the slots of any deleted locks. When a lock is deleted or a new lock is created, every instance of the GUI running anywhere on the network is notified to update its display.

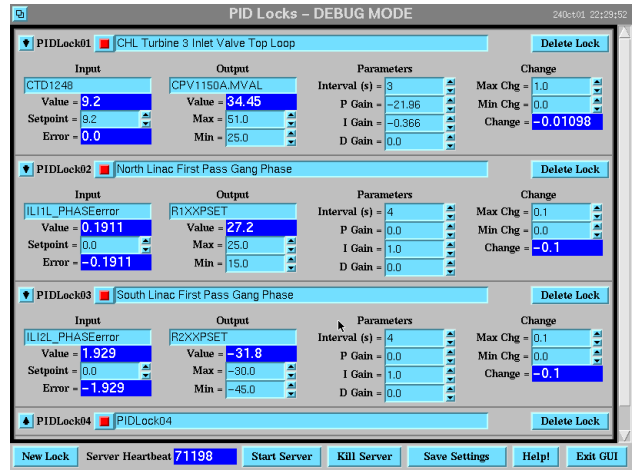

**Figure 2: The PID Lock GUI**

## 4 FUTURE DIRECTIONS

### 4.1 Additional Lock Types

The remaining slow lock types in use at Jefferson Lab will be integrated into the Generic Lock Server architecture. For this to happen the architecture must be extended to allow MIMO lock types. It must also allow for locks that must be calibrated at runtime like the Orbit Locks and locks that use the on-line accelerator model server for calibration like the Energy Locks.

### 4.2 Server Security

Access to all CEBAF IOC's is controlled through CA security. To function, the Generic Lock Server must run with a user ID that is allowed to write to the IOC's. However, because it is designed to be flexible, the CDEV Generic Server has no built-in security model. This means that any user connected to the accelerator network can write to the virtual attributes of the lock devices and create a lock that writes to any channel in the system, effectively circumventing CA security. Plugging this loophole is a top development priority. A security layer will be added to the lock server allowing writes only by specific groups of users from specific machines. In the longer term, it would be useful if the attributes of a lock would inherit the security of the lock's output channel.

### 4.3 PID Auto-tuning

Determining appropriate gains for PID controllers is a non-trivial task. To make the PID Lock class more usable by non-experts it would be very desirable to have a means to automatically characterize the closed loop transfer function and determine appropriate PID gains for optimum stability, if they exist. Many such algorithms exist [9]. If one can be found that is sufficiently general, it might be integrated into the lock server architecture or could exist as a separate process, making it more useful for tuning EPICS CPID records as well.

### 4.4 Dynamic Linking

The very low degree of coupling among the various lock and server classes means that it is not necessary to statically link the lock classes with a server. Following the model that CDEV uses with its service classes, the code for individual lock types could be dynamically loaded as needed. This would allow completely new types of locks to be added to a running server without even restarting it, much less rebuilding it.

### 4.5 Expanding the Lock Namespace

One limitation imposed by CDEV is that all device names must be declared in static Device Definition Language (DDL) files that are used by clients to map device names to the appropriate services and servers. Thus, although new locks can be created in the server with arbitrary names, clients cannot find them unless they use names pre-declared in a common DDL file. This prohibits the use of descriptive device names for dynamically created locks. A dedicated client like the Generic Lock GUI could be designed to dynamically discover the lock names, but they would still not be available to general purpose clients like archivers.

CDEV can be configured to fall through to a particular service when a device name is not found in the DDL file. For many sites that use CDEV with EPICS, this default is configured to use CA. The Portable CA Server (PCAS) [10] suggests a way to make fully dynamic virtual devices. A PCAS could be set up to host the virtual device/attribute pairs needed by such devices.

## 5 ACKNOWLEDGEMENTS